\documentclass[twocolumn,twoside,slac_two]{revtex4}
\usepackage{graphicx}
\usepackage{fancyhdr}
\usepackage{epstopdf}
\begin{document}

\title{Recent Results from IceCube and AMANDA}


\author{T. DeYoung}
\affiliation{Department of Physics, Pennsylvania State
  University, University Park, PA 16802, USA}
\author{for the IceCube Collaboration}

\begin{abstract}
  IceCube is a cubic kilometer neutrino telescope under construction
  at the South Pole, a successor to the first-generation AMANDA
  telescope.  IceCube is now three quarters complete, with completion
  expected in early 2011, and data taken with the partially built
  detector already provides a sensitivity surpassing the complete
  AMANDA-II data set.  Results from searches for astrophysical sources
  of neutrinos and for evidence of dark matter with both AMANDA and
  IceCube are summarized.  We also discuss plans for Deep Core, an
  enhancement of IceCube designed to extend its sensitivity to
  neutrinos below the TeV scale.
\end{abstract}

\maketitle

\thispagestyle{fancy}


%%%%%%%%%%%%%%%%%%%%%%%%%%%%%%%%%%
\section{Introduction}
Nearly a century after the discovery of the cosmic rays, considerable
uncertainty remains regarding the sources accelerating these
particles.  It is believed that cosmic rays at the highest energies, above the
``ankle'' in the cosmic ray spectrum around $10^{19}$ eV, are
accelerated by extragalactic objects such as gamma ray
bursts (GRBs) or active galactic nuclei (AGN).
At lower energies, the sources of the cosmic rays are likely galactic
objects such as supernova remnants (SNRs), although to date there is
no conclusive evidence for cosmic ray acceleration by any of these
objects.

The IceCube neutrino telescope, now under construction at the South
Pole, attempts to detect very high energy (VHE, roughly TeV to PeV
scale) neutrino emission from the accelerators of the cosmic rays.
Neutrino emission is expected from such objects, since the highly
energetic environments required to accelerate the cosmic rays will
likely contain matter or radiation fields with which the accelerated
hadrons may interact before escaping the source.  These interactions
will produce charged $\pi$ and $K$ mesons which decay to muon and
electron neutrinos (and antineutrinos, which we will not distinguish
in these proceedings) \cite{Learned:2000sw}.  The neutrinos 
oscillate over distance between their sources and the Earth,
leading to a general expectation of flavor equality at detection,
although this is not guaranteed: deviations from flavor equality could
provide a variety of information regarding the source environment or
probe exotic fundamental physics.
%\cite{Rachen:1998fd,Kashti:2005qa,Anchordoqui:2003vc,Razzaque:2005ds,Beacom:2003nh}.

At Earth, VHE neutrinos which undergo charged or neutral current
interactions with a nucleon can be detected via the Cherenkov
radiation emitted by secondary particles produced in the highly
inelastic interaction.  Neutrino flavor identification is possible
based on the topology of the Cherenkov radiation: $\nu_\mu$ charged
current interactions produce a highly energetic muon that travels for
hundreds of meters or kilometers on a straight trajectory, leading to
long straight tracks in the detector.  Electron neutrinos and neutral
current interactions of all neutrinos produce localized showers of
particles, leading to an approximately spherical emission of Cherenkov
light.  Tau neutrinos can produce a variety of signatures, depending
on their energy and decay mode
\cite{Learned:1994wg,Beacom:2003nh,DeYoung:2006fg}.

\begin{figure}[h]
  \centering
  \includegraphics[width=80mm]{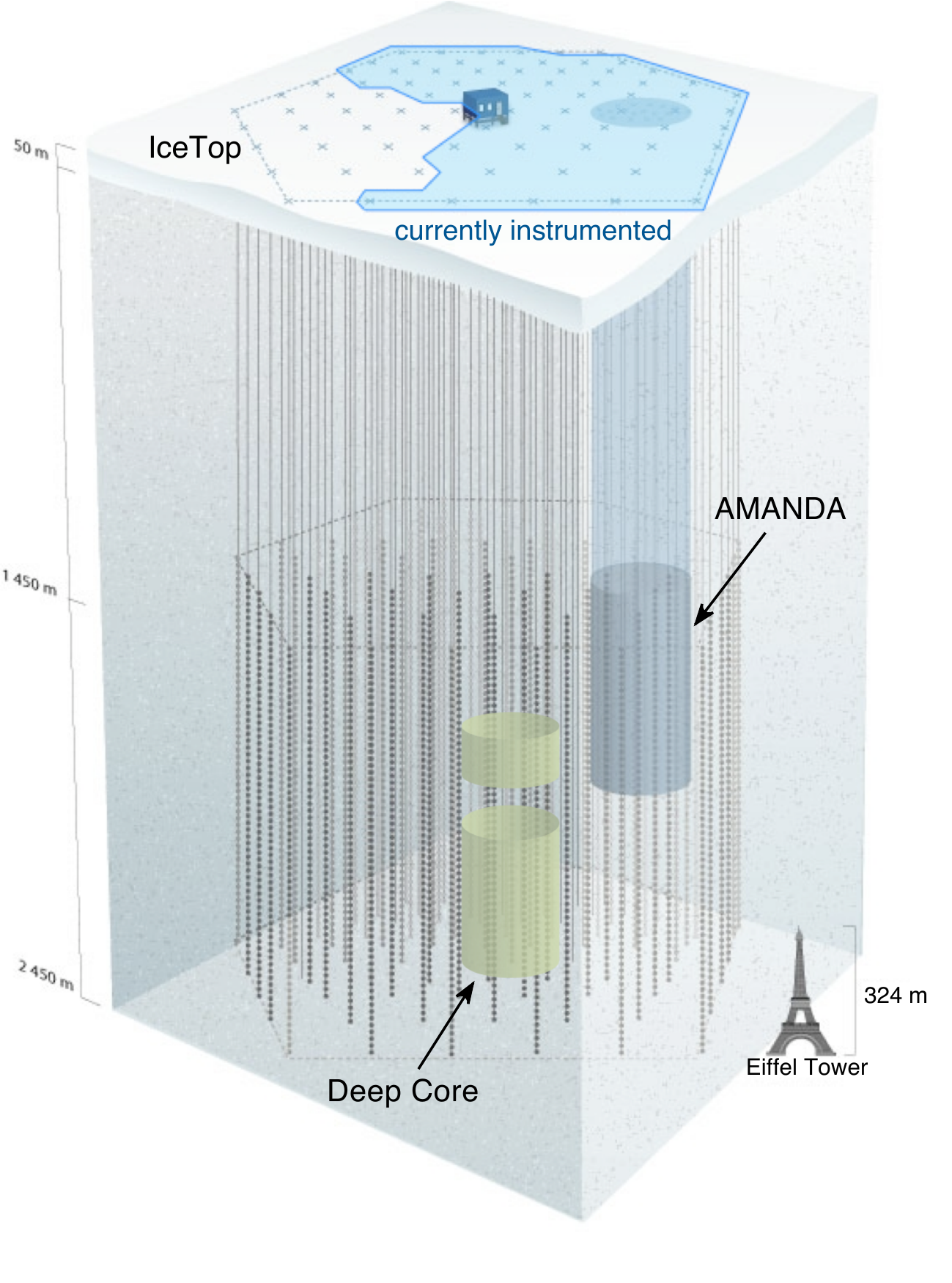}
  \caption{Schematic of the IceCube detector, including the initial
    AMANDA array, the new Deep Core low energy extension, and the
    IceTop air shower array.  The portion of the detector installed as
    of 2009 is shown in blue, and the Eiffel Tower is shown for
    scale.} \label{fig:layout} 
\end{figure}

The IceCube telescope uses a cubic kilometer of the Antarctic ice cap
as the Cherenkov medium to detect neutrinos, instrumenting the ice
with a three dimensional array of photomultiplier tubes (PMTs) housed
in digital optical modules (DOMs).  The completed detector will
consist of 5160 DOMs attached to 86 vertical strings at depths between
1450 m and 2450 m below the surface of the ice cap.  In addition, an
extensive air shower array on the surface, known as IceTop, will
detect cosmic rays.  The layout of the array is shown in
Fig.~\ref{fig:layout}.  At present, 59 strings are installed
and operational, with the remaining 27 strings to be deployed in the
coming two austral summers.  Also shown in Fig.~\ref{fig:layout} is
the AMANDA telescope, which operated as an
independent instrument from 2000-2006 and was integrated into IceCube
from 2007-2008.  AMANDA consisted of 687 optical modules on 19
strings, with the bulk of the detector at depths of 1500 to 2000 m.
The AMANDA optical modules were simpler than their IceCube
counterparts, with considerably smaller dynamic range and less
ability to resolve individual photoelectrons.  In 2009, AMANDA was
decommissioned and the first string of its replacement, Deep Core, was
deployed.

%%%%%%%%%%%%%%%%%%%%%%%%%%%%%%%%%%
\section{Atmospheric Neutrinos}

\begin{figure}[h]
  \centering
  \includegraphics[width=80mm]{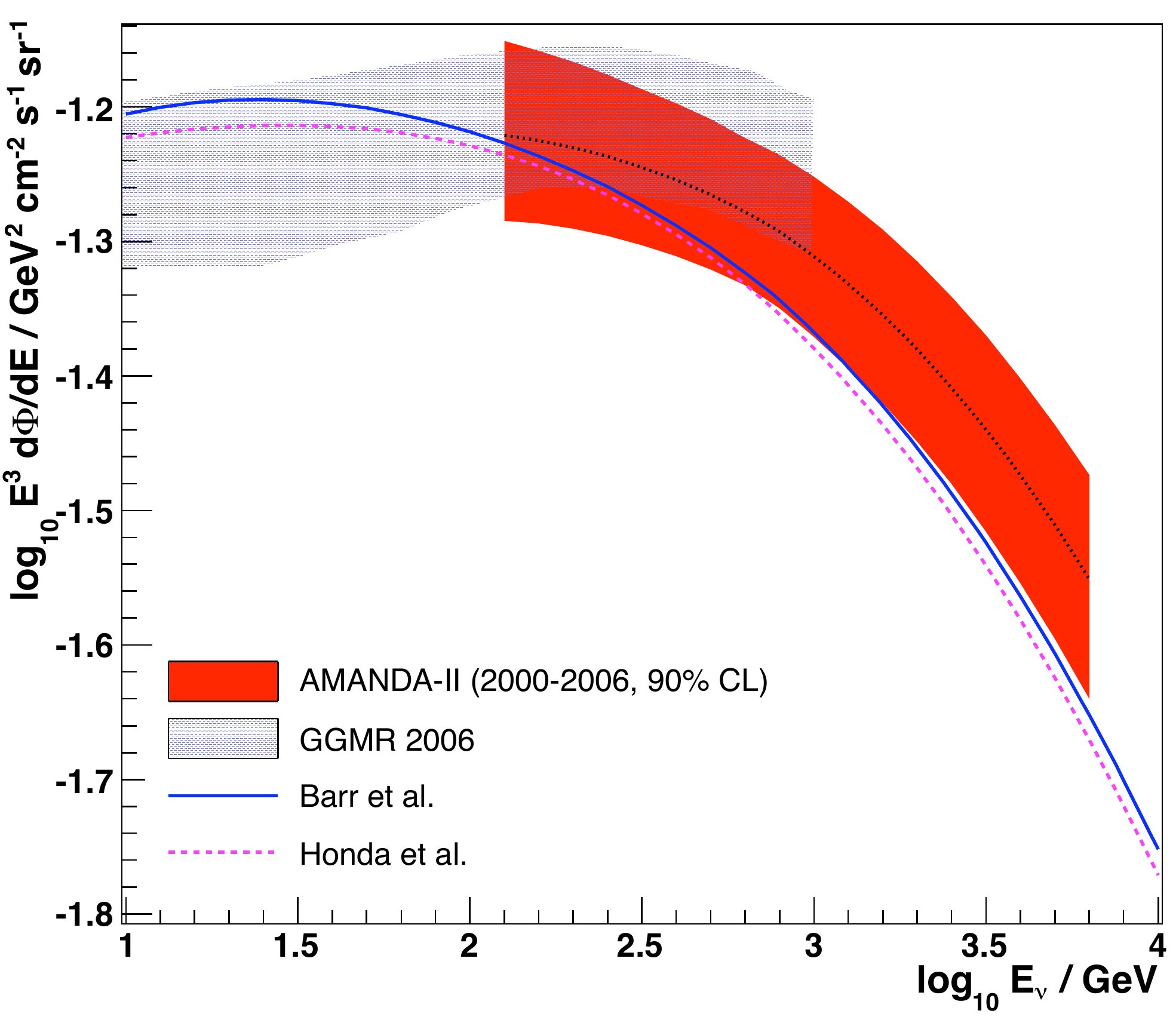}
  \caption{AMANDA measurement of the atmospheric $\nu_\mu +
    \bar{\nu}_\mu$ flux \cite{Abbasi:2009nf}, compared to theoretical
    models (lines) and a measurement (GGMR) extracted from the
    Super-Kamiokande data \cite{GonzalezGarcia:2006ay}.  The flux
    measured by AMANDA is slightly larger and harder than the Bartol
    calculation \cite{Barr:2004br}, but consistent at the 99\% C.L. or
    better in each parameter.  The total theoretical uncertainty is
    larger than the spread between the curves would suggest, due to
    uncertainties in measured parameters common to both calculations.
  } \label{fig:atmflux}
\end{figure}

The bulk of the neutrinos detected by IceCube are not produced in
astrophysical accelerators, but rather in cosmic ray air showers in
the Earth's atmosphere.  These atmospheric neutrinos constitute a
diffuse background to the astrophysical neutrinos from the cosmic ray
sources, but they also provide a useful calibration beam for the
detector.  It is also possible to use the high-statistics atmospheric
neutrino data set to make a variety of fundamental physics
measurements~\cite{Abbasi:2009nf}.  Figure~\ref{fig:atmflux} shows the
measurement of the atmospheric $\nu_\mu$ spectrum extracted from the
AMANDA seven year data set, as compared to two theoretical
calculations \cite{Barr:2004br,Honda:2006qj}.  The spread between the
two calculations does not indicate the full theoretical uncertainty,
as both models rely on physics inputs which are themselves uncertain.
AMANDA observes a spectrum which is slightly harder and slightly more
numerous than the models, although consistent with the central values
of the calculations at the 90\% C.L. in normalization and the 99\%
C.L. in spectral index.

%%%%%%%%%%%%%%%%%%%%%%%%%%%%%%%%%%
\section{Point Sources of Neutrinos}

\begin{figure*}[t]
  \centering
  \includegraphics[width=120mm]{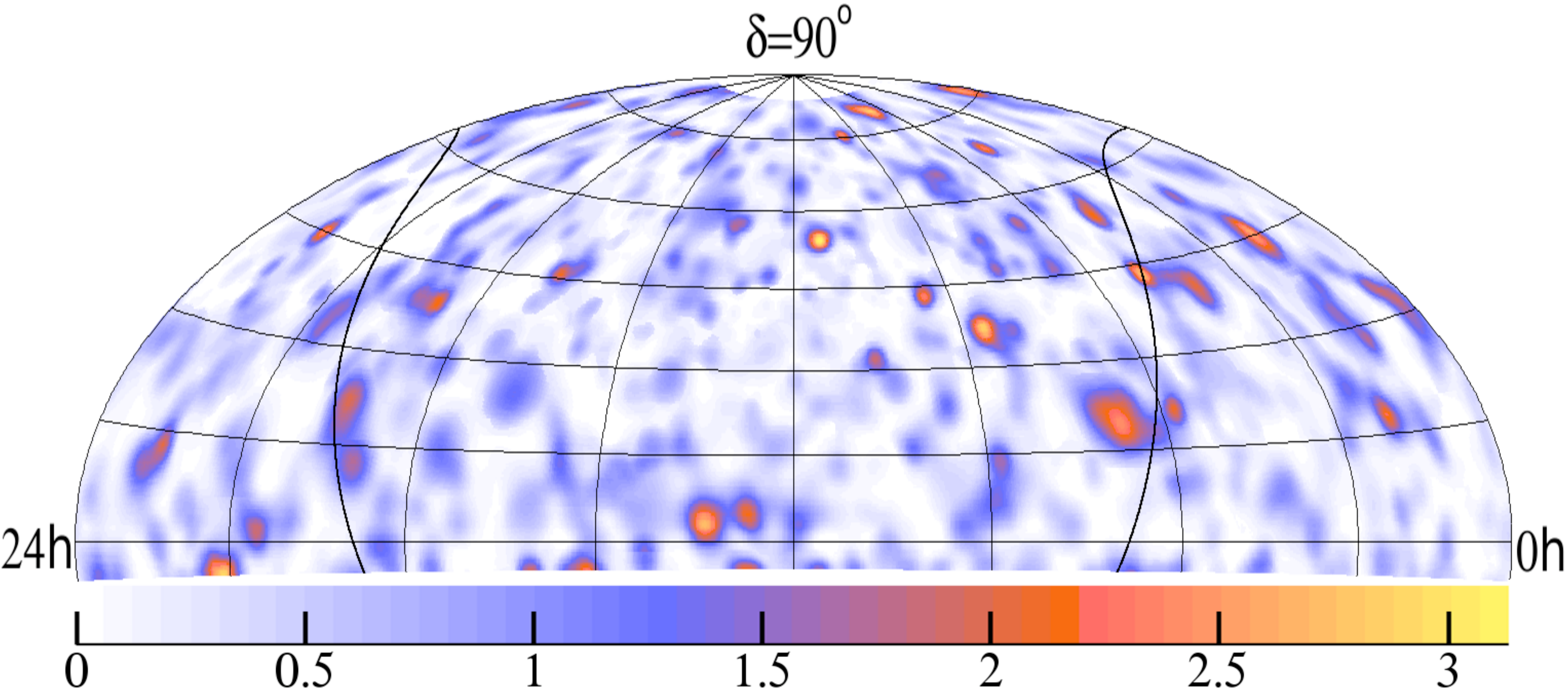}
  \caption{Map of the significances of deviations from the background
    in the AMANDA 2000-06 point source search.  The scale converts
    these significances to standard deviations $\sigma$ in the normal
    distribution.  The most significant deviation is equivalent to
    3.38$\sigma$ before accounting for the number of points on the
    sky; a deviation at least this significant is found in 95\% of
    scrambled (signal-free) sky maps, indicating that this observation
    is consistent with the background.  The thin line indicates the
    Galactic plane.} \label{fig:amanda_ptsrc}
\end{figure*}

The primary goal of IceCube is to detect individual sources of
astrophysical neutrinos produced by cosmic ray accelerators.  Such a
source would be identifiable as a localized excess on the sky,
normally with a harder energy spectrum than the atmospheric neutrino
background.
We search for sources using an unbinned (likelihood) approach.  
% At any
% point on the sky, a likelihood
% \[
% \mathcal{L} = \prod_{i=1}^N \left(\frac{n_s}{N} \mathcal{S}_i + (1 -
%   \frac{n_s}{N}) \mathcal{B}_i\right)
% \]
% can be calculated using all events within a $\pm8^\circ$ declination
% band, assuming $n_s$ of them are associated with a source at the test
% position and
% the remaining $N - n_s$ are atmospheric background events.  The
% probability density functions $\mathcal{S}$ and $\mathcal{B}$ describe
% the chances of nearby events being signal or background
% neutrinos, respectively.  The signal pdf accounts for the angular
% resolution of the detector and the spectral index $\gamma$ of
% the hypothetical neutrino source, with the event energy estimated
% using the number of channels hit.  The background pdf includes an
% energy term using the known atmospheric spectrum but no angular
% resolution, since the atmospheric flux is essentially isotropic.
%
% At every point on the sky, the likelihood ratio 
% \[
%  \lambda = -2 \log\left(\frac{\mathcal{L}(n_s =
%      0)}{\mathcal{L}(\hat{n}_s,\hat{\gamma})}\right) 
% \]
% is calculated, comparing the best fit number of neutrinos $\hat{n}_s$
% associated with a source at that position and the best fit spectral
% index $\hat{\gamma}$ of the source to the null hypothesis.  
The true
significance of deviations from the null hypothesis is assessed by
generating an ensemble of background maps, using the real data set
with every event scrambled in right ascension.
Figure~\ref{fig:amanda_ptsrc} shows the signficances of each point in
the Northern Hemisphere sky produced by AMANDA in seven years of
operation \cite{Abbasi:2008ih}.  This data set includes 6,595 neutrino
candidate events collected over 3.8 years of exposure (after
accounting for readout dead time between events and summer maintenance
periods).  The most significant point in the map has a pre-trials
significance equivalent to 3.38$\sigma$; a comparably significant
point is found in 95\% of scrambled background maps, indicating that
the map is consistent with the absence of any neutrino
point source.

\begin{figure*}[t]
  \centering
  \includegraphics[width=150mm]{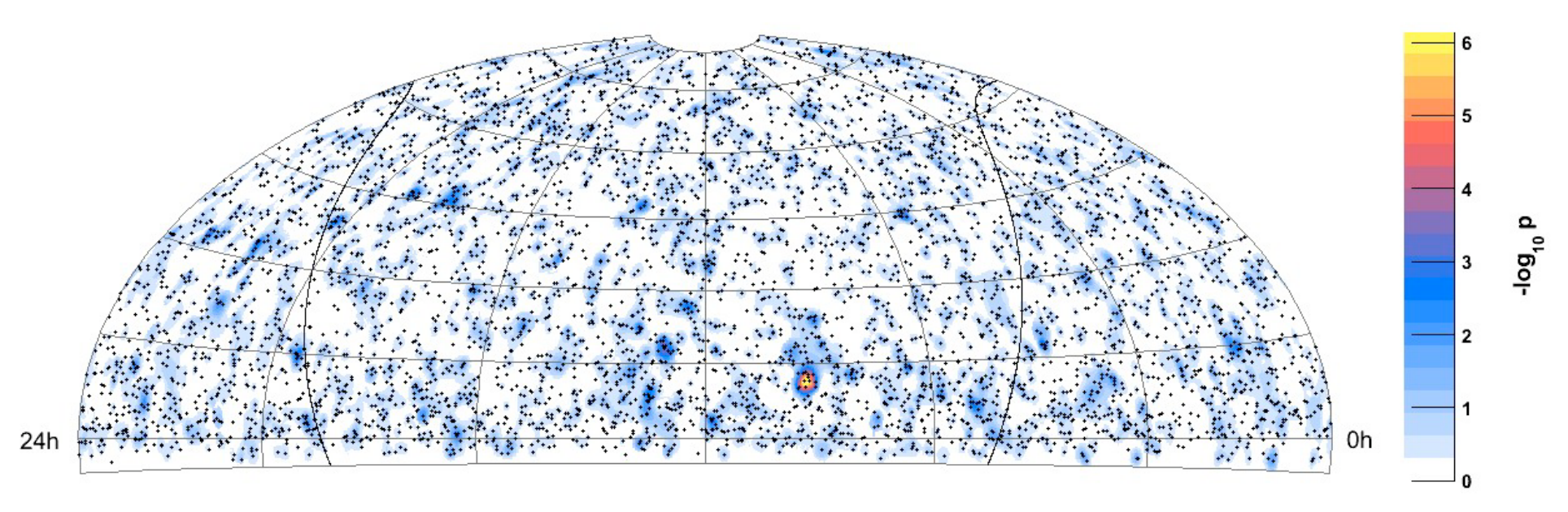}
  \caption{Map of significances on the northern sky in the 22-string
    2007 IceCube point source search.  Black dots indicate individual
    neutrino events, and the color scale indicates the significance of
    the likelihood ratio observed at each point.  The most significant
    point on the sky has a significance ($p$-value) of $7 \times
    10^{-7}$; there is a 1.34\% chance of observing such a deviation
    under the background hypothesis, so this observation is consistent
    with background.
  } \label{fig:ic22_ptsrc}
\end{figure*}

A similar search was undertaken using the 2007 IceCube data set,
containing 276 days of data from 22 strings taken between May 2007 and
April 2008 \cite{Abbasi:2009iv}.  A total of 5,114 neutrino candidate
events was observed, consistent with an expectation of 4600 $\pm$ 1400
atmospheric neutrino events and 400 $\pm$ 200 cosmic ray muons
misreconstructed as upgoing neutrino events.  The same likelihood
method was used to analyze the data, with the exception that the
angular uncertainty was estimated for each neutrino event based on the
width of the optimum in the likelihood space used to reconstruct the
direction of the event, rather than using a generical overall angular
resolution for all events.  The resulting map of significances is
shown in Figure~\ref{fig:ic22_ptsrc}.  The most significant point in
this map has a pre-trials significance of $7 \times 10^{-7}$; using
randomized sky maps and also accounting for the fact that an
independent search for emission from a list of candidate sources was
also conducted using this data set, the probability of observing such
a deviation from the null hypothesis is estimated at 1.34\%,
consistent with a background fluctuation.

\begin{figure}[h]
  \centering
  \includegraphics[width=80mm]{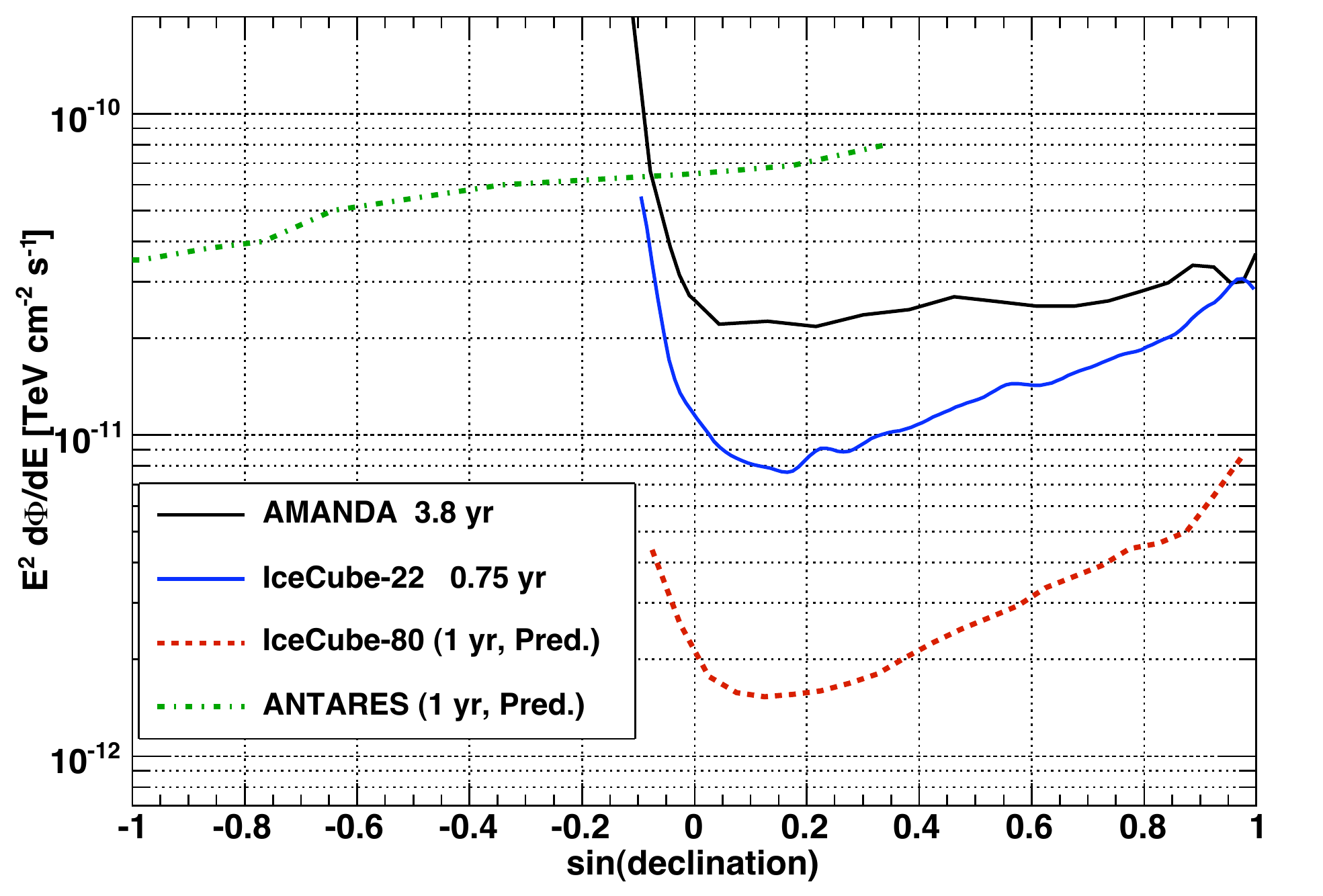}
  \caption{Average 90\% C.L. upper limit on the $\nu_\mu +
    \bar{\nu}_\mu$ flux (assuming flavor equality) from a point source
    as a function of declination, compared to predicted sensitivities
    for IceCube and ANTARES.} \label{fig:ptsrc_sens}
\end{figure}

The sensitivity of these searches is shown in
Figure~\ref{fig:ptsrc_sens}, as a function of declination.  The
sensitivity of the IC22 search is significantly better than that of
the AMANDA search, despite being based on a much shorter exposure
(approximately 9 months of data collected in one year of operation,
compared to 3.8 years of data collected in 7 years with AMANDA).  The
instantaneous sensitivity of IceCube, even when only one quarter
built, is significantly better than that of AMANDA.  The complete
detector will be even more sensitive, a factor of five better than the
present analysis with one year of operation.

A list of 26 potential neutrino emitters, known through their
electromagnetic emissions, was drawn up {\it a priori}, including AGN,
SNRs, and TeV gamma ray sources identified by Milagro in the Cygnus
region \cite{Abdo:2007ad} (now identified with GeV pulsars or pulsar
wind nebulae \cite{Abdo:2009ku} on the basis of correlations with the
Fermi Bright Source List \cite{Abdo:2009mg}).  The likelihood ratio, as
defined above, was calculated for each of these sources using the
AMANDA data set.  Because there are many fewer locations of interest,
this restricted source is less subject to statistical trials penalties
than the full-sky search, and is thus somewhat more sensitive.  A
similar analysis was performed on the IC22 data, using an expanded
candidate list of 28 sources.  A selection of the results from these
searches is shown in Table~\ref{tab:ptsrc}, including those with the smallest
$p$-values in each search.  In neither case are these $p$-values
inconsistent with the background hypothesis: one expects to obtain $p
\leq 0.0086$ for at least one of 26 sources in 20\% of signal-free sky
maps, and $p \leq 0.071$ in 66\%.  The 90\%
C. L. upper limits placed on $\nu_\mu$ emission from the sources,
assuming $E^{-2}$ spectra and $1:1:1$ flavor ratios, are
also shown in Table~\ref{tab:ptsrc}.

\begin{table*}[t]
  \begin{center}
    \caption{Selected $p$-values and 90\% C. L.~upper limits on
      $\nu_\mu + \bar{\nu}_\mu$ fluxes $E_\nu^2 \, dN/dE \leq
      \Phi_{90} \times 10^{-12}$ TeV/cm$^2$ s, from searches for
      neutrino emission from predefined candidate sources with AMANDA
      and IceCube-22.  Dashes indicate that the $p$-value was not
      calculated because the best-fit number of signal events was
      zero.}
    \begin{tabular}{|l|r|r|r|r|r|r|}
      \hline 
      \textbf{Source} & \textbf{decl. [$^\circ$]} & \textbf{r.a. [h]}
      & \textbf{AMANDA $p$-value} & \textbf{IC22 $p$-value} 
      & \textbf{$\Phi_{90} (\textrm{AM})$} & \textbf{$\Phi_{90} (\textrm{IC22})$} \\
      \hline 
      Crab Nebula        & 22.01 &   5.58 & 0.10     & $-$    & 46.4 & 10.35  \\
      Geminga             & 17.77 &   6.57 & 0.0086 & $-$    & 63.9 & \hphantom{0}9.67  \\
      MGRO J2019+37 & 36.83 & 20.32 & 0.077   & 0.25   & 48.4 & 25.23  \\
      LS I +61 303       & 61.23 &   2.68 & 0.034   & $-$    & 73.7 & 22.00  \\
      XTE J1118+480   & 48.04 & 11.30 & 0.50     & 0.082 & 25.9 & 40.62 \\
      Cygnus X-1        & 35.20 & 19.97 & 0.57     & $-$    & 20.0 & 14.60  \\
      Mrk 421              & 38.21 & 11.07 & 0.82     & $-$    & 12.7 & 14.35  \\
      Mrk 501              & 39.76 & 16.90 & 0.22     & $-$    & 36.4 & 14.44 \\
      1ES 1959+650    & 65.15 & 20.00 & 0.44     & 0.071 & 33.8 & 59.00  \\
      M87                    & 12.39 & 12.51 & 0.43     & $-$    & 22.5 & \hphantom{0}7.91  \\
      \hline
    \end{tabular}
    \label{tab:ptsrc}
  \end{center}
\end{table*}

\section{Neutrinos from Gamma Ray Bursts}

Searches for neutrino emission from gamma ray bursts were also
conducted, using both the AMANDA and IceCube data sets.  Because
external detectors such as X-ray satellites detect the GRB, the
IceCube search can be tailored carefully to the specific times and
locations on the sky, leading to a drastically lower background rate.
The expected event rates are also low, typically considerably less than one
event per burst, so the searches are normally conducted by `stacking'
a collection of bursts and searching for neutrino emission from the
ensemble as a whole.

\begin{figure}[h]
  \centering
  \includegraphics[width=80mm]{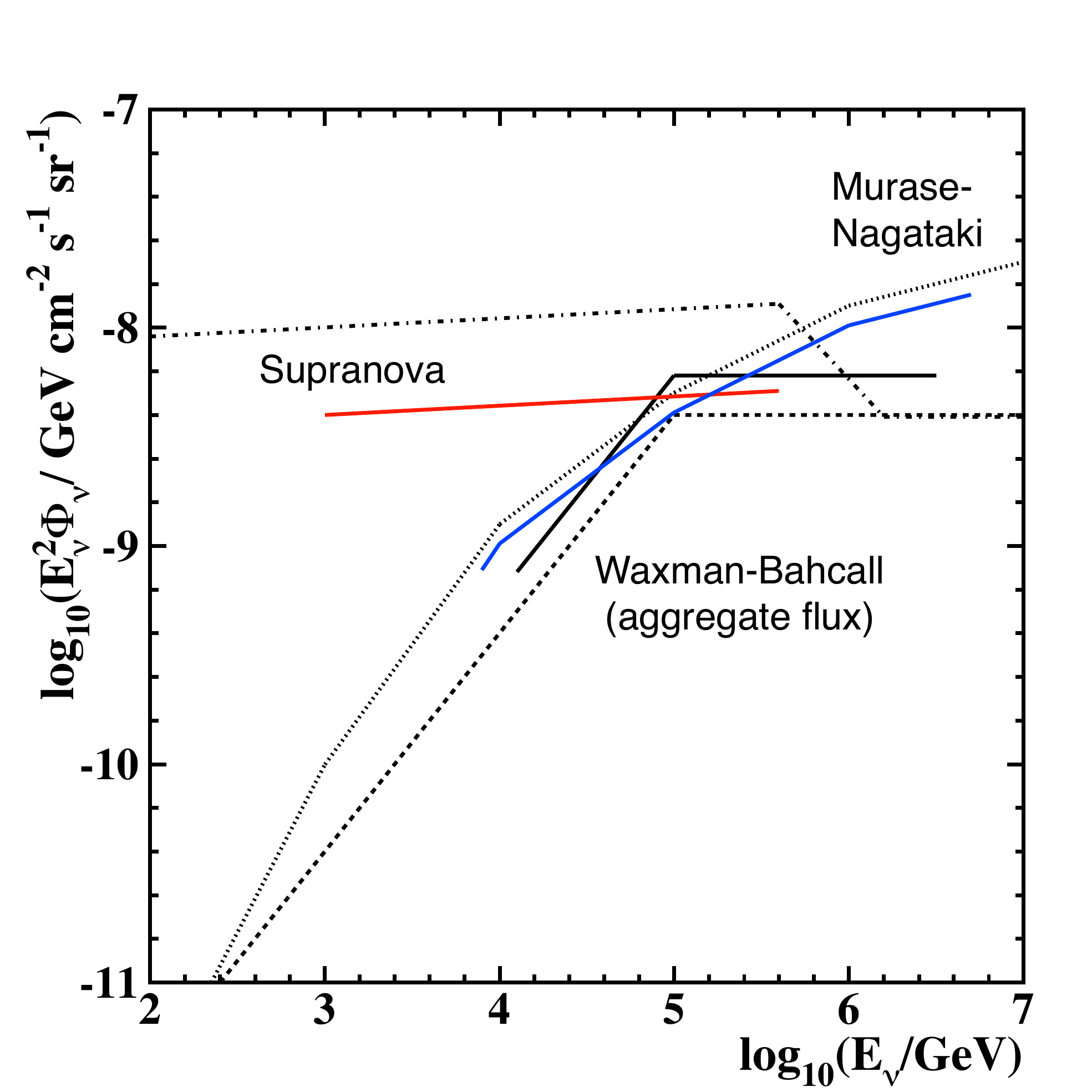}
  \caption{Integral limits at 90\% C.L. on several $\nu_\mu +
    \bar{\nu}_\mu$ flux models (details in text) using AMANDA
    observations of 419 GRBs.  Dashed lines indicate the model
    predictions, while solid lines show the relative level of the
    integral limit and the energy ranges of sensitivity.  The energy
    ranges indicated contain 90\% of the expected flux.  The flux
    limits are for the full sky ($4\pi$ sr), although only bursts from
    the Northern Hemisphere were used in the
    analysis.} \label{fig:amanda_grb}
\end{figure}

The result of searches for emission from 419 GRBs observed in the
Northern Hemisphere during stable AMANDA operations between 1997 and
2003 \cite{Achterberg:2007nx} is shown in Fig.~\ref{fig:amanda_grb}.
Three theoretical models are shown: the Waxman-Bahcall
\cite{Waxman:2002wp} (divided by 2 to account for
neutrino oscillations) and Murase-Nagataki \cite{Murase:2005hy}
calculations based on the assumption that GRBs are the sources of the
ultrahigh-energy cosmic rays, and a `supranova'
model~\cite{Razzaque:2002kb} assuming that all GRBs are preceded by
supernovae which produce ideal circumburst environments for neutrino
production (only 60 GRB observations were used in placing this limit).

Because limits are placed on the integrated flux predicted by the
models, our constraints are given in terms of ``model rejection
factors'' (MRFs), essentially the scaling factor at which the model would be
just ruled out at 90\% confidence.  MRFs less than 1
indicate that the model is excluded at the stated confidence level.  The
MRFs for the three models are 1.36 for the Waxman-Bahcall model, 0.92
for the Murase-Nagataki parameter set A, and 0.45 for the supranova model under the
assumptions mentioned above.  

\begin{figure}[h]
  \centering
  \includegraphics[width=80mm]{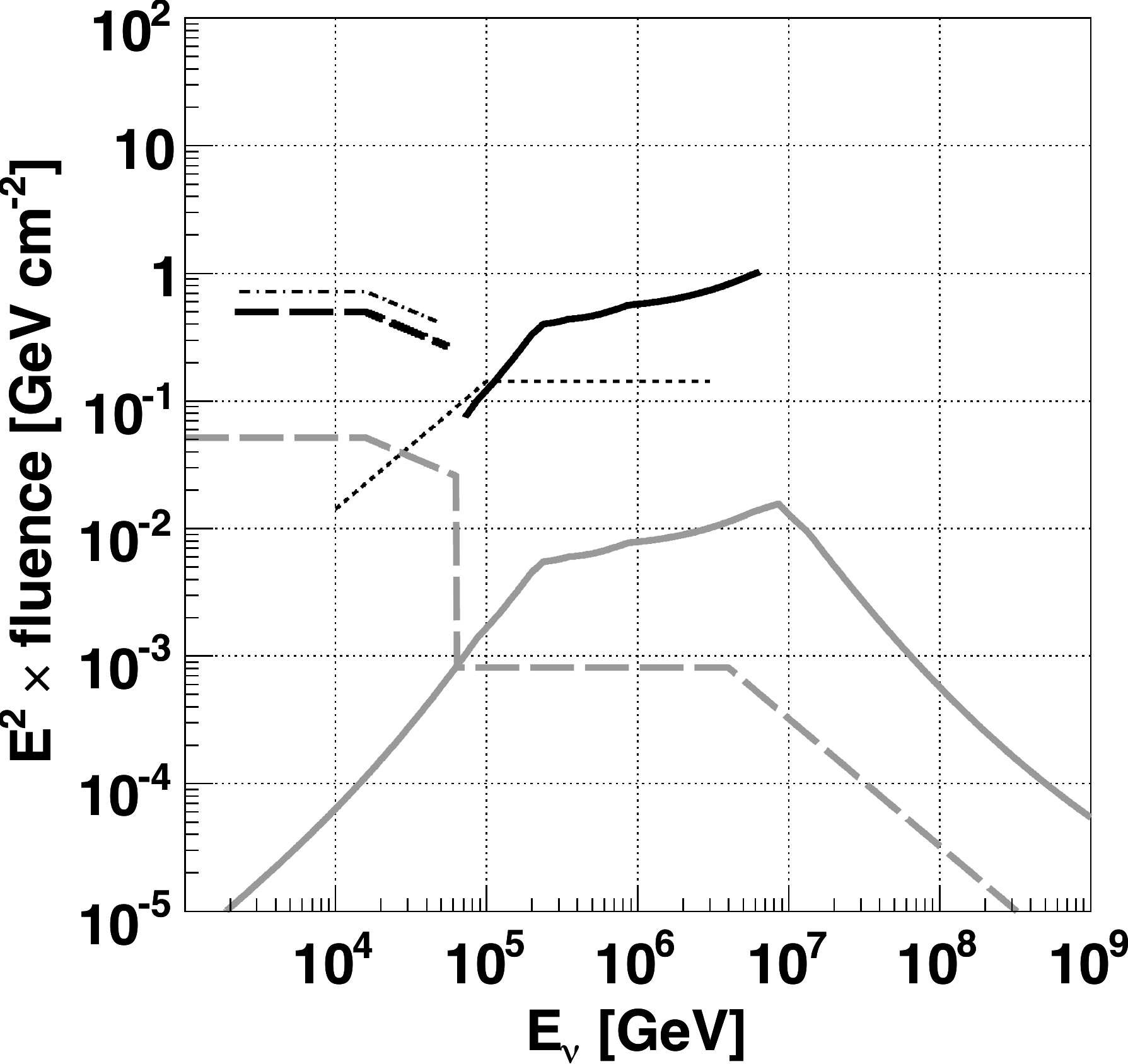}
  \caption{Integral limits (heavy black lines) at 90\% C. L. on
    several predictions of total $\nu_\mu + \bar{\nu}_\mu$ fluence
    based on IceCube observations of 41 bursts with the 22 string 2007
    configuration.  The gray lines indicate the precursor model
    \cite{Razzaque:2003uv}, peaked at lower energies, and our
    prediction for the prompt fluence (following \cite{Guetta:2003wi})
    at higher energies, using the observed characteristics of
    individual bursts in a Waxman-Bahcall like model.  The solid lines
    show the relative level of the integral limits on these fluences,
    with the energy ranges indicated containing 90\% of the expected
    fluence.  The thin dotted lines are the AMANDA limits 
    % on the precursor model and the standard W-B GRB flux for
    %comparison.
    from Fig.~\ref{fig:amanda_grb} converted to fluences for comparison.
  } \label{fig:icecube_grb}
\end{figure}

The results of a comparable search using IceCube observations of 41
GRBs using the 22 string 2007 data set \cite{Abbasi:2009ig} are shown
in Figure~\ref{fig:icecube_grb}.  Limits are placed on a
model similar to the Waxman-Bahcall GRB prediction but taking into account the
characteristics of individual bursts following the method of \cite{Guetta:2003wi}, as well as
the precursor model of Razzaque et al.~\cite{Razzaque:2003uv}. 

\section{Diffuse Astrophysical Neutrino Fluxes}

In addition to individual sources, one can also search for a diffuse
flux of astrophysical neutrinos from an ensemble of sources too faint
to resolve individually.  Any such flux must be separated from the
nearly isotropic flux of atmospheric neutrinos using the fact that the
atmospheric spectrum is quite soft ($dN/dE \sim E^{-3.7}$) while the
spectra of astrophysical sources are generically much harder ($E^{-2}$
for ideal shock acceleration).

\begin{figure*}[t]
  \centering
  \includegraphics[width=160mm]{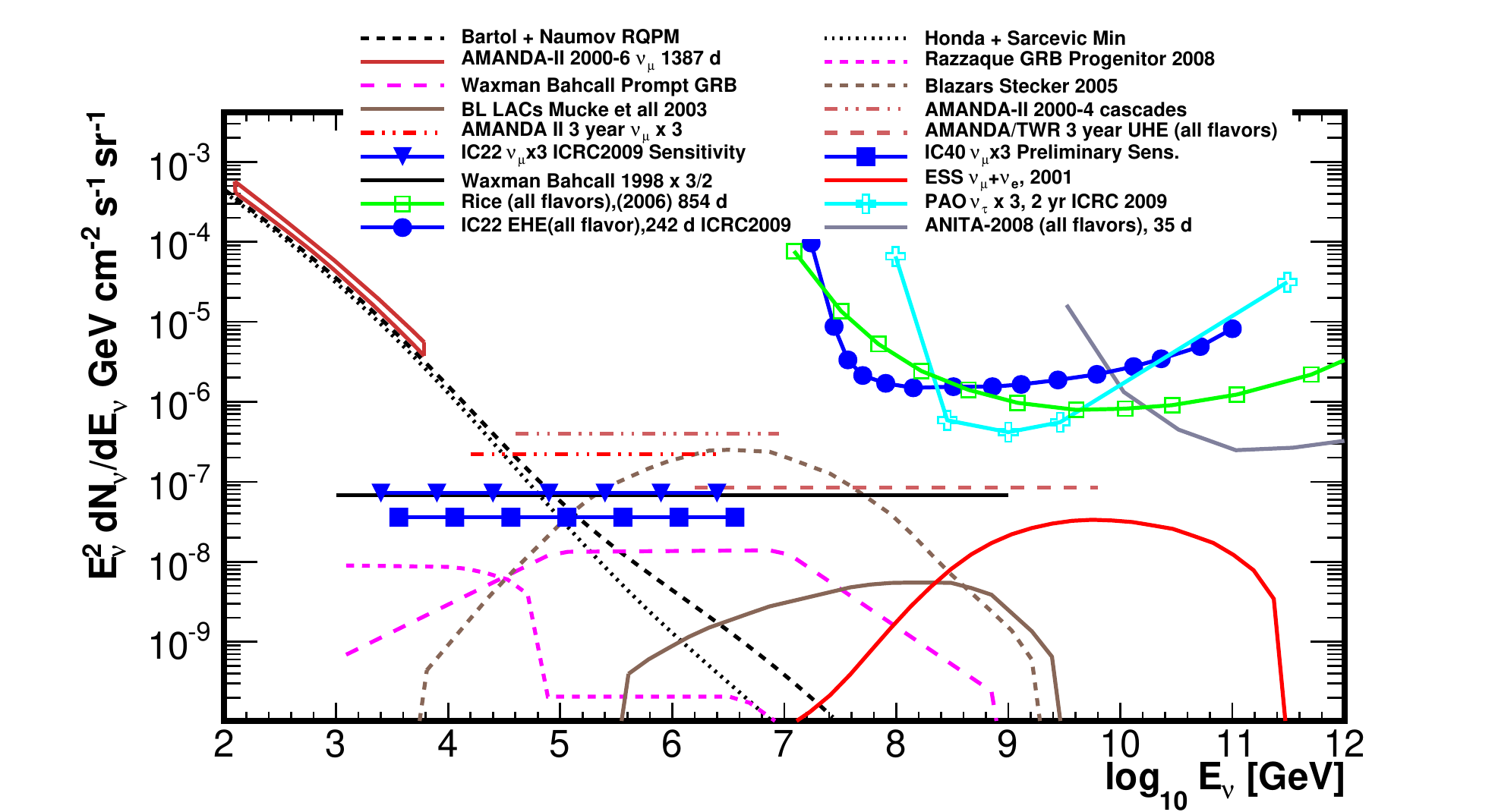}
  \caption{Limits on diffuse astrophysical neutrino fluxes from AMANDA
    and IceCube, compared with theoretical models and limits from
    other experiments.  The AMANDA atmospheric muon neutrino
    measurement (Fig.~\ref{fig:atmflux}) is the red region at upper
    left.  Limits on possible $E^{-2}$ diffuse fluxes from AMANDA,
    Baikal, and IceCube are shown as horizontal lines; these are
    integral limits on such fluxes, with the energy range from which
    the central 90\% of the events would be expected shown.  Fluxes
    which surpass those limits for only a small part of that energy
    range would not be excluded by these limits.  Except for the
    atmospheric neutrinos, all fluxes include all flavors and have
    been rescaled if necessary assuming flavor equality at Earth.  At
    upper right, a number of differential limits from ultrahigh energy
    neutrino experiments are shown; the levels of these limits are not
    directly comparable to the levels of the integral
    limits.} \label{fig:diffuse}
\end{figure*}

Two independent searches for diffuse fluxes were undertaken using
AMANDA data. The first \cite{Achterberg:2007qp} was closely related to
the standard muon neutrino analysis used to search for point sources
of neutrinos, with cuts optimized for higher energy neutrinos and with
a cut placed on the reconstructed energy of the neutrino candidate
events. This search used data from 2000--2003, and the results are
shown in Fig.~\ref{fig:diffuse}. As a benchmark, the limit placed
on a hypothetical diffuse $\nu_\mu$ flux $dN/dE = \Phi_0 E^{-2}$ at
90\% C. L. is $\Phi_0 \geq 7.4 \times 10^{-8}$ GeV cm$^{-2}$ s$^{-1}$
sr$^{-1}$. For such a flux, 90\% of the signal neutrinos would have
had energies between 16 TeV and 2.5 PeV, which are the bounds of limit
shown in Fig.~\ref{fig:diffuse}. The limit is placed on the
integrated flux, and so cannot be directly compared to specific
theoretical models; instead, the predicted signal for each model must
be simulated to find the model rejection factor (MRF). Limits on several
specific models are also shown (as thin lines parallel to the
predicted fluxes) in Fig.~\ref{fig:diffuse}, and appear in
Table~\ref{tab:diffuse}. In addition to astrophysical models, a flux
corresponding to the upper bound on generic optically thin
($\tau_{n\gamma} < 1$) pion photoproduction
sites \cite{Mannheim:1998wp} was tested. The upper limit from the
analysis is a factor of 0.22 of the MPR bound over the region from 10
TeV to 630 TeV.

\begin{table*}[t]
  \begin{center}
    \caption{Limits on several theoretical models of diffuse muon
      neutrino fluxes from the two AMANDA analyses.  The number of
      events that would have been detected $n_{sig}$ and the ``model
      rejection factor'' (MRF), the ratio of the upper limit to the
      predicted flux, are shown.  MRFs less than 1 indicate that the
      model is excluded at the 90\% C. L.}
    \begin{tabular}{|l|r|r|r|r|l|}
      \hline 
      \textbf{Source} & \textbf{$n_{sig} (\textrm{HE})$} & \textbf{MRF
        (HE)} & \textbf{$n_{sig} (\textrm{UHE})$} & \textbf{MRF
        (UHE)} & \textbf{Model} \\  
      \hline 
      Active Galactic Nuclei          
      & 1.7   & 1.6   &   1.8 &   2.9 & Stecker \cite{Stecker:2005hn} \\
      & 1.4   & 2.0   &   5.9 &   0.9 & MPR \cite{Mannheim:1998wp} \\
      &         &         &   8.8 &   0.6 & Halzen \& Zas \cite{Halzen:1997hw} \\
      &         &         & 20.6 &   0.3 & Protheroe \cite{Protheroe:1996uu} \\
      &         &         &   0.3 & 18.0 & Mannheim \cite{Mannheim:1995mm} RL A \\
      &         &         &   4.5 &   1.2 & Mannheim \cite{Mannheim:1995mm} RL B \\
      \hline
      Starburst Galaxies
      & 1.1   & 21.1 &         &         & Loeb \& Waxman \cite{Loeb:2006dk} \\ 
      \hline
      Prompt Atmospheric $\nu_\mu$
      & 0.4   & 60.3 &         &         & MRS GBW \cite{Martin:2003us} \\
      & 4.7   &   5.2 &         &         & Naumov RQPM \cite{Naumov:2002dm} \\
      & 16.1 &   1.5 &         &         & Zas et al.~\cite{Zas:1992ci} Charm C \\
      & 26.2 & 0.95 &         &         & Zas et al.~\cite{Zas:1992ci} Charm D \\
      \hline
    \end{tabular}
    \label{tab:diffuse}
  \end{center}
\end{table*}

A second analysis based on data from 2000--2002 exploited
idiosyncracies of the hardware response to the extremely bright events
produced by ultrahigh energy (UHE) neutrinos,
such as afterpulsing in the PMTs, to extend the range of the detector
to much higher energies \cite{Ackermann:2007km}.  The limit for the
$E^{-2}$ benchmark $\nu_\mu$ flux is $9.0 \times 10^{-8}$ GeV
cm$^{-2}$ s$^{-1}$ sr$^{-1}$ at 90\% C. L. (assuming a $1:1:1$ flavor
ratio), over a range from $2 \times 10^5$ GeV to $10^9$ GeV, as shown
in Fig.~\ref{fig:diffuse}. It should be noted that the analysis
was sensitive to all flavors of neutrinos; for a $E^{-2}$ spectrum
with flavor equality, the flavor ratio of the detected events would
have been approximately $2:2:1$.  Limits on specific theoretical models
are calculated separately for each model and are omitted
from Fig.~\ref{fig:diffuse} for clarity but are shown in
Table~\ref{tab:diffuse}.  

Searches for diffuse astrophysical neutrinos are also underway using
data from the 22-string 2007 configuration of IceCube (IC22) and the
40-string 2008 configuration (IC40).  The IC22 sensitivity of $7.5
\times 10^{-8}$ GeV cm$^{-2}$ s$^{-1}$ sr$^{-1}$ is very nearly at the
Waxman-Bahcall level, the benchmark diffuse  flux
level obtained by normalizing the parent proton population to the observed
cosmic ray flux.  The expected sensitivity from the IC40 data is well
below the Waxman-Bahcall flux, indicating that the completed detector
will probe astrophysically significant flux levels.

\section{Search for Dark Matter}

In addition to neutrinos produced during the acceleration of cosmic
rays, IceCube would be sensitive to neutrinos produced in the decay of
dark matter particles.  For example, weakly interacting dark matter
particles (WIMPs) such as neutralinos could scatter off nucleons in
the Sun and become trapped in the solar gravitational well, where they
could produce high energy neutrinos through various annihilation
channels.  Such a search is highly complementary to direct dark matter
searches using heavy elements such as germanium or xenon, which seek
to take advantage of coherent scattering of the WIMP off the nucleons
in the atom.  If the nucleon-WIMP scattering is independent of the
nucleon spin (SI), then this coherence will increase the cross section
$\sigma_{SI}$ by a factor of the square of the atomic mass of the
target.  However, if the primary coupling is spin-dependent (SD),
coherence is lost and the cross section $\sigma_{SD}$ will be much
smaller.  WIMP capture in the Sun, which is primarily made of light
nuclei, provides a useful probe of models where the scattering is
primarily spin-dependent \cite{Halzen:2005ar}.

It should be noted that, as opposed to direct dark matter detection
experiments, in this indirect approach assumptions must be made
regarding the annihilation products, which will affect the energy
spectrum of the neutrinos emerging from such processes.  The hardest
spectrum would come from annihilations to $W^+W^-$ ($\tau^+\tau^-$ for
the lowest masses) and the softest
from annihilations to $b\bar{b}$, so as limiting cases we assume the
WIMPs decay exclusively to those particles.  The actual annihilation
cross section is irrelevant so long as the Sun is old enough for the
WIMP population to have reached equilibrium, with captures balancing
annihilations.  

\begin{figure}[h]
  \centering
  \includegraphics[width=80mm]{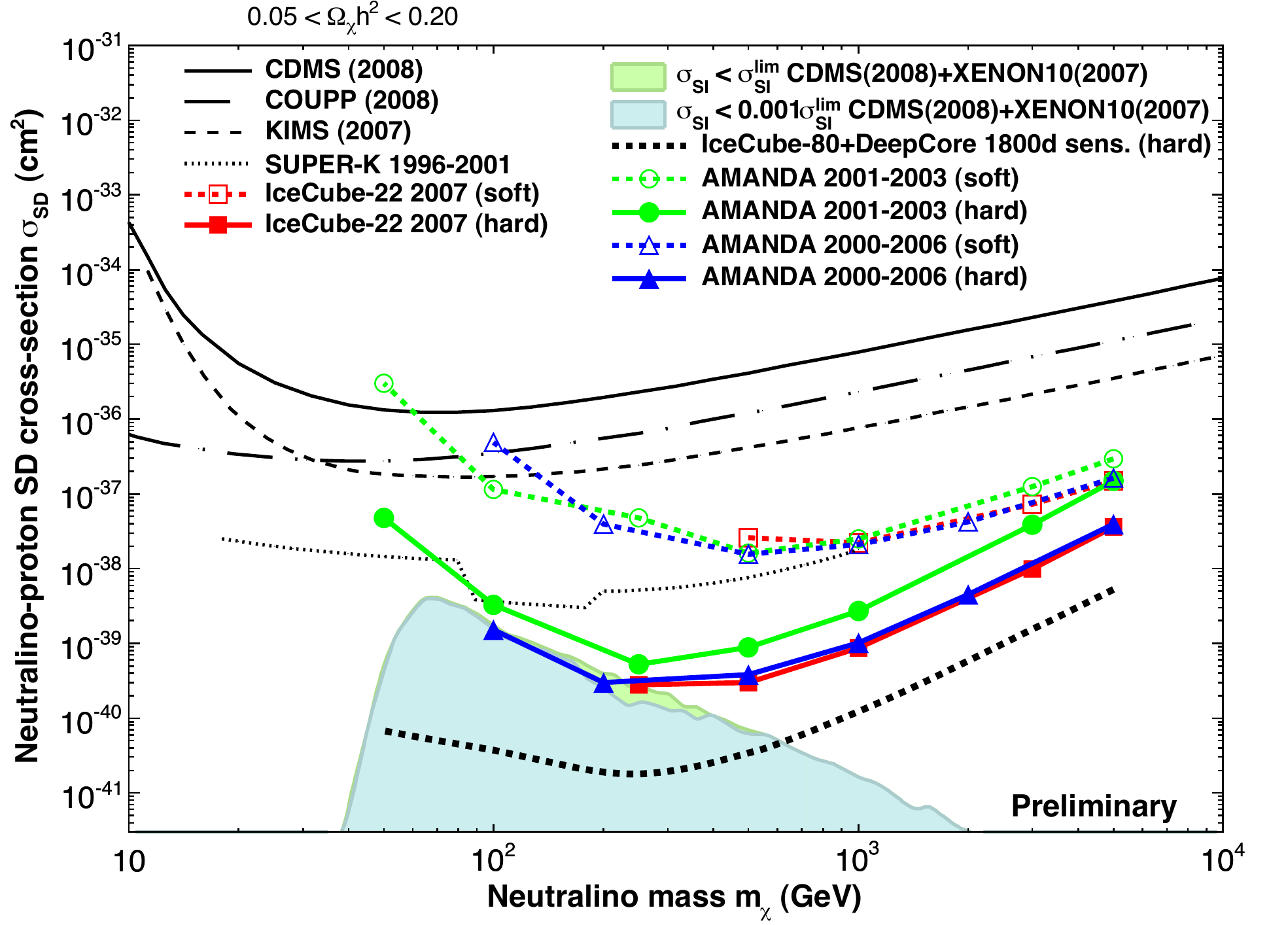}
  \caption{Limits on the spin-dependent neutralino-proton cross
    section derived by searching for high energy neutrinos from WIMP
    annihilation in the Sun.  The limits are shown for the extreme
    cases of hard ($\chi \chi \rightarrow W^+W^-$) and soft
    ($b\bar{b}$) neutrino spectra.  The shaded region indicates the
    MSSM parameter space allowed by existing limits on the
    corresponding spin-independent cross section predicted for that
    combination of SUSY parameters.  The thin green shaded region
    would be ruled out if existing limits on $\sigma_{SI}$ are
    improved by a factor of $10^3$.} \label{fig:wimps}
\end{figure}

Searches for high energy neutrinos from the Sun have been undertaken
with both AMANDA \cite{Ackermann:2005fr} and IC22
\cite{Abbasi:2009uz}.  The limits on the SD scattering cross section
produced by these searches are shown in Figure~\ref{fig:wimps}, in the
limiting cases of hard and soft annihilation spectra.  Limits from
Super-Kamiokande \cite{Desai:2004pq} and a number of direct search
experiments
\cite{Angle:2007uj,Ahmed:2008eu,Lee.:2007qn,Behnke:2008zza} are also
shown, as well as the allowed MSSM neutralino parameter space based on
direct detection limits on the SI cross section.  Tighter SI limits
would rule out only a limited region of the allowed space, so IceCube
provides a useful counterpart to future direct detection experiments.

In the future, the low energy response of the IceCube detector will be
greatly augmented by the addition of the Deep Core low energy
extension, visible in Fig.~\ref{fig:layout}.  Deep Core will comprise the seven
innermost standard IceCube strings, as well as six new strings
deployed in a ring of radius 72 m around the central string. The six
new strings will each mount 60 DOMs, 50 of which will be deployed on a
7 m spacing between 2100 m and 2450 m below the surface. The remaining
10 DOMs will be deployed at shallower depths to improve the efficiency
of detection of extremely vertical background muons. With a radius of
125 m and a height of 350 m, the instrumented volume of Deep Core will
instrument 15 Mton of ice, with expected sensitivity to neutrinos at
energies as low as $\sim$10 GeV.

The new DOMs will be identical to standard IceCube DOMs except that
they will use a new model of PMT developed by Hamamatsu to increase
the quantum efficiency of the photocathode.  Lab tests with assembled
DOMs indicate the sensitivity of the high-QE PMTs is approximately
30\% higher than that of standard DOMs.  The denser DOM spacing and
higher DOM sensitivity combined will increase the collection of
photons in the Deep Core volume by approximately a factor of 5.
Furthermore, the ice at Deep Core depths is significantly more
transparent than that at shallower depths, with optical attenuation
lengths of 40--45 m compared to 20--25 m in the top of the detector.
The significantly improved light collection in Deep Core
translates to much higher sensitivity to relatively dim, low energy
neutrino events.  Additionally, the
bulk of IceCube can be used to detect and veto atmospheric muons
penetrating to Deep Core. The ratio of the atmospheric muon trigger
rate to the atmospheric neutrino trigger rate in IceCube is
approximately $10^6$; initial Monte Carlo studies 
indicate that veto efficiencies on this order are achievable with
relatively good signal efficiency.  

The first of the six new Deep Core strings was successfully deployed
at South Pole in the 2008-09 austral summer.  Preliminary evaluations
of the performance of the hardware {\it in situ} confirm expectations
from laboratory studies. 
Deployment of the remaining five new strings, as well as the
standard strings that compose the Deep Core array, is scheduled to be
complete by February 2010.  Deep Core will enable study of a number of
topics in neutrino physics and searches for neutrinos from point
sources in the southern sky, including the Galactic center region.  It
will also greatly improve the sensitivity of IceCube to low-mass
WIMPs.  A preliminary sensitivity curve for Deep Core using 1800 days
of data is shown in Fig.~\ref{fig:wimps}.

\section{Outlook}

IceCube construction is proceeding well, with
completion expected in 2011.  IceCube will be augmented
with the Deep Core array, to be completed in 2010, which will
extend its capabilities to energies as low as 10 GeV.
Initial results from the partially built detector, including only one
quarter of the final array, are already providing sensitivities beyond
those of the complete seven-year AMANDA-II data set,
and this sensitivity will improve rapidly as construction progresses.
Within a few years the sensitivity of IceCube will be sufficient to
probe astrophysical neutrino fluxes below the Waxman-Bahcall level.
In addition, IceCube and Deep Core will permit indirect searches for dark
matter well beyond existing limits.

\begin{acknowledgments}
  We acknowledge the support from the following agencies:
  U.S. National Science Foundation-Office of Polar Program,
  U.S. National Science Foundation-Physics Division, University of
  Wisconsin Alumni Research Foundation, U.S. Department of Energy, and
  National Energy Research Scientific Computing Center, the Louisiana
  Optical Network Initiative (LONI) grid computing resources; Swedish
  Research Council, Swedish Polar Research Secretariat, and Knut and
  Alice Wallenberg Foundation, Sweden; German Ministry for Education
  and Research (BMBF), Deutsche Forschungsgemeinschaft (DFG), Germany;
  Fund for Scientific Research (FNRS-FWO), Flanders Institute to
  encourage scientific and technological research in industry (IWT),
  Belgian Federal Science Policy Office (Belspo); the Netherlands
  Organisation for Scientific Research (NWO); M. Ribordy acknowledges
  the support of the SNF (Switzerland); A. Kappes and A. Gro\ss{}
  acknowledge support by the EU Marie Curie OIF Program.
\end{acknowledgments}

%\begin{thebibliography}{9}   % Use for  1-9  references

\end{document}